\documentclass[12pt]{article}
\newtheorem{theorem}{Theorem}
\newtheorem{lemma}{Lemma}

\newtheorem{example}{Example}

\newtheorem{note}{Note}

\begin{document}

\begin{center}
{\Large \bf On Computational Aspects of the  Fourier-Mukai
Transform}
\end{center}
\smallskip

\begin{center}
{\bf Nikolaj M. Glazunov} \end{center}
\smallskip
\begin{center}
{\rm Glushkov Institute of Cybernetics NAS \\
 03187 Ukraine Kiev GSP 680 Glushkov prospekt 40 \\
 Email:} {\it glanm@yahoo.com }
\end{center} \smallskip

 \begin{center} {\bf  Abstract} \end{center}
{\it We survey and investigate some computational aspects of the
Fourier-Mukai transform.}
\smallskip

\section{Introduction}
 Let $f(x)$ be a function $f:{\bf Z}/n{\bf Z} \rightarrow {\bf C},
 \; n \in {\bf N}$.
 Then its Fourier transform from the
 time domain into its frequency domain is given by the discrete
 Fourier transform
\[
{\hat f}(\omega) = \sum_{x = 0}^{n - 1} f(x) \exp(2\pi i
\frac{\omega x }{n}).
\]
Here $\omega x$ is the perfect pairing (in the case the scalar
product) on the product of the time and of the frequency domains.
It is well known also that the solution of a linear differential
equation with constant coefficients is related by the Fourier
transform to a solution of the polynomial equation. The
Fourier-Mukai transform is a strong generalization of mentioned
and some another approaches.  Let $A$ be an abelian variety,
${\hat A}$ its dual abelian variety and ${\mathcal P}$ the
Poincar{\'e} divisor on $A \times {\hat A}$. Let $D^{b}(A)$ and
$D^{b}({\hat A})$  be derived categories of bounded complexes of
sheaves on $A$ and ${\hat A}$ respectively. A Fourier-Mukai
transform was defined by Mukai as an exact equivalence
\[
{\mathcal F}{\mathcal M}: \; D^{b}(A) \rightarrow D^{b}({\hat A})
\]
between derived categories of above mentioned bounded
complexes\cite{Glazunov:mukai,Glazunov:mukaia}.  For this
transform analogies of the Fourier Inversion Theorem and the
Parseval Theorem are valid. Works by B. Bartocci, U. Bruzzo, D.
Ruip{\'e}rez\cite{Glazunov:bartocci&bruzzo&hernandez} and A.
Maciocia\cite{Glazunov:maciocia} have generalized this approach to
another classes of sheaves and varieties. Now by a Fourier-Mukai
transform $$ {\mathcal F}{\mathcal M}: \;  D^{b}(Y) \rightarrow
D^{b}(X)$$ an exact equivalence between bounded derived categories
of coherent sheaves on two smooth projective varieties $X$ and $Y$
is understood. It is known that the derived categories of coherent
sheaves on some projective varieties are equivalent to the derived
categories of representations of $n-$vertices quivers. For
instance, the derived category of coherent sheaves on ${\bf
P^{2}}$ is equivalent to the derived category of representations
of $3-$vertices quivers with relations\cite{Glazunov:bondal}.
There is a very interesting connection between Fourier-Mukai
transforms and mirror symmetry\cite{Glazunov:thomas}. A quiver can
be interpreted as a directed
graph\cite{Glazunov:draxler&norenberg}. We investigate some
computational aspects of Fourier-Mukai transforms. This paper is a
continuation of\cite{Glazunov:glazunovms,Glazunov:glazunovga}.\\
 The organization of the article is as follows. In Section 2 we
 recall from the computational point of view some facts about Abelian
 varieties. In Section 3 we very shortly consider parameterization of
 Abelian varieties and some moduli spaces. In Section 4 we give
 follow to Mukai the definition of the Fourier-Mukai transform.
In Section 5 we present our computer algebraic method for
computation of cutsets of some quivers and its implementation.

\section{Abelian varieties}

   Let ${\bf C}^g$ be a complex vector space of the dimension $g$ and
$\Lambda$ a lattice in ${\bf C}^g$ with a bases
$\{{\bf a}_{1}, \ldots ,{\bf a}_{2g}\}$. For a lattice vector
${\bf a} = (a_{1}, \ldots , a_{2g}) \in \Lambda$ its length is
denoted by $|{\bf a}|$.
The factor ${\bf C}^{g}/\Lambda = A$ is the commutative compact
topological group (complex torus). Recall the case $g = 1.$
\begin{example}
Elliptic curves. \\ Let $\Lambda = \{{\bf a}, {\bf b}\}, \; {\bf
a}, {\bf b} \in {\bf C}, \; {\bf a}/{\bf b} \notin {\bf R}$. Then
$E = {\bf C}/\Lambda$ is an elliptic curve. In the case every such
lattice defines an elliptic curve. What is the simplest
representation of $\Lambda$? \\
\begin{lemma}
Basis reduction in $\Lambda$. \\ In any such lattice $\Lambda$
there exists (under the rotations) one and only one basis with
conditions:  \\ (i) $|{\bf b}| \ge |{\bf a}|$;            \\ (ii)
length of the projection ${\bf b}$ on ${\bf a} \; \le \frac12
|{\bf a}|$; \\ (iii) the angle$({\bf a},{\bf b}) < \pi/2$ (acute
angle). \\ So we can reduce a basis of $\Lambda$ to the form
$\Lambda = \{1, \tau\},$ where $Im \: \tau > 0, \; -\frac12 \le Re
\: \tau \le \frac12, \; |\tau| \ge 1$.
\end{lemma}
Let
\[
h = \left| \begin{array}{cc}
a&b\\
c&d
\end{array}\right|, \; \det h = 1, \; a, b, c, d \in {\bf Z}.
\]
Two lattices $\Lambda = \{1, \tau\}, \; \Lambda^{'} = \{1, \tau^{'}\}$
are equivalent if $\tau^{'} = \frac{a\tau + b}{c\tau + d}.$
In the case there is a birational isomorphism between
elliptic curves $E = {\bf C} / \Lambda$ and $E^{'} = {\bf C} / \Lambda^{'}$.
\end{example}
If $g > 1$ then there are groups $A = {\bf C}^{g}/\Lambda$
which are not Abelian varieties\cite{Glazunov:mumford}. The complex torus
$A$ is an Abelian variety if and only if there exist ${\bf R}-$bilinear
antisymmetric form $F(x,y)$ such that the form $F(x,ix)$ is a positive
definite Hermitian form that takes integer values at points of $\Lambda$.
The form $F(x,y)$ is called the {\em polarization} on $\Lambda$
and a pair $(\Lambda, F)$ is called the polarized Abelian variety.
For computational purposes a good idea is  to reduce the bases of
the lattice to a some simple form.
It is well known the bases reduction of $\Lambda$ in the case $g = 1$
 (Example 1).
\subsection{On basis reduction in lattices  (the case $g > 1$)}
  Let a lattice $\Lambda = \{{\bf a}_{1}, \ldots ,{\bf a}_{2g}\},
 \; {\bf a}_{i} = (a_{i1}, \ldots , a_{i2g}), $
has coordinates with the condition $a_{ij} \in {\bf Q}.$ \\ There
are {\em Minimum basis problem},  {\em Shortest lattice vector
problem} and {\em LLL lattice
reduction}\cite{Glazunov:cassels,Glazunov:grotschel&schryver&lovasz}.\\
The method of a solution of the Minimum basis problem is based on
the theory of successive minima, developed by
Minkowski\cite{Glazunov:cassels}. By the method we try to find
short linear independent vectors one after one. The solution of
the  Shortest lattice vector problem is based on Minkowski's
theorem on convex body\cite{Glazunov:cassels}. Probably that both
the methods are $NP-$hard. \\ In some applications the concept
{\em reduces} means something like nearly orthogonal. The $L^3$
algorithm\cite{Glazunov:lenstra&lenstra&lovasz} accept as input
any basis ${\bf a}_{1}, \ldots ,{\bf a}_{2g}$ of $\Lambda$, and it
gives a reduced basis ${\bf b}_{1}, \ldots ,{\bf b}_{2g}$ of
$\Lambda$. In general, a lattice may have more then one reduced
basis. The ordering of the basis vectors is not arbitrary. The
algorithm has a very good theoretical complexity (polynomial-time
in the length of the input parameters).

\section{Parameterization of Abelian varieties and moduli spaces}

 Let $(\Lambda, F)$ be a polarized Abelian variety.
Denote by $M^{tr}$ the transposition of a matrix $M.$  There is a
canonical form of Abelian varieties. In the form the lattice of
the Abelian variety has the representation
\[
  \Omega = (E_{g},Z),
\]
 where $E_{g}$ is the identity matrix and $Z = X + iY$ is a
complex $g \times g$ matrix with conditions: a) $Z = Z^{tr}, \;
 X = X^{tr}, \; Y = Y^{tr}$ ($X, \; Y$ are reals); b) $Y > 0$ is the
matrix of a positive quadratic form. The set $\{Z\} = {\bf H}_{g}$
of the matrices is called the {\em Siegel upper half-space}. Let
\[
J
= \left| \begin{array}{cc} 0&E_{g}\\ -E_{g}&0
\end{array}\right|
$$ be the symplectic matrix. Let $$ M = \left| \begin{array}{cc}
A&B\\ C&D
\end{array}\right| , \; M \in GL(2g,{\bf Z}), \; M^{tr}JM = J,
\]
be the modular group. The modular group acts on ${\bf H}_{g}$ by
the formula $$ Z' = (AZ + B)(CZ + D)^{-1}. $$ If matrices $X$ and
$Y$ have integer coefficients, then all data with the exception of
complex the unity $i$ are defined over ${\bf Z}.$ Therefore they
and their factors ${\mathit mod} p$ for prime $p$ are well-adapted
for computers. Moduli
spaces~\cite{Glazunov:mamford,Glazunov:harris&morrison} are used
for specification and investigation of classes of objects which
could be algebraic curves (or, more generally, schemes), sheaves,
vector bundles, morphisms and others.
Recall that the algebraic curve over complex numbers could be
viewed as a Riemann surface with branches. What is moduli?
Classically Riemann claimed that $3g - 3$ (complex) parameters
could be for Riemann surface of genus $g$ which would determine
its conformal structure (for elliptic curves, when $g = 1,$ it is
needs one parameter).
 Parameter varieties is a class of moduli spaces. These varieties
are very convenient tool for computer algebra investigation of
objects that are parameterized by the parameter varieties. If we
investigate a class of algebraic varieties over integers, then
 the first step of the investigation in many cases is the analysis
of the varieties over prime finite fields.  We have used the
approach for investigation of rational points on families of
hyperelliptic curves~\cite{Glazunov:glazunov}. Expansion of the
group law of an Abelian variety near $0$ defines a formal group of
the dimension one in the case of elliptic
curves~\cite{Glazunov:tate} and of the dimension  $ g > 1$ for
$g-$dimensional Abelian varieties~\cite{Glazunov:honda}. Results
about formal group over finite fields and over $\rho$-adic
fields~\cite{Glazunov:glazunovfg} can be useful for investigation
of the groups over ${\bf Z}.$ The paper~\cite{Glazunov:kornyak}
demonstrates computational advantages of the localization under
computation of cohomologies.

\section{Elements of category theory and the Fourier-Mukai transform}
\subsection{Equivalence of Categories}
  Let ${\mathcal C}$ be a category\cite{Glazunov:maclane},
  $Ob \; {\mathcal C}$ its class
  of objects, and for $a, \; b \in Ob \; {\mathcal C}, \;{\mathcal
  C}(a,b)$ the class of morphisms (arrows) from $a$ to $b.$ An
  arrow $u: a \rightarrow b$ is {\it the equivalence} in ${\mathcal C}$ if
there is
  an arrow $u': b \rightarrow a$ such that $u'u = 1_{a}$ and $uu' =
  1_{b}$. A ${\it functor} \; {\mathcal F}$ from a category ${\mathcal C}$ to
a category ${\mathcal K}$ is a function which maps $ Ob({\mathcal
C}) \rightarrow Ob({\mathcal K}),$ and which for each pair $a, b$
of objects of ${\mathcal C}$ maps $ {\mathcal C}(a,b) \rightarrow
{\mathcal K}({\mathcal F}(a),{\mathcal F}(b)),$ while satisfying
the two conditions: \\
 ${\mathcal F}id_{a} = id_{{\mathcal F}a}$ for every
$a \in Ob({\mathcal C}),$ \\
 ${\mathcal F}(fg) = {\mathcal F}(f) {\mathcal F}(g).$      \\

Let ${\mathcal Funct}({\mathcal C},{\mathcal D})$ be
  the category of functors from ${\mathcal C}$ to ${\mathcal D}$ with
  natural transformations as morphisms. An equivalence in
  ${\mathcal Funct}({\mathcal C},{\mathcal D})$ is {\it the
  equivalence} between categories ${\mathcal C}$ and ${\mathcal
  D}$.
  The following theorem is a particular case of the theorem
  from\cite{Glazunov:maclane}.
  \begin{theorem}
 A functor ${\mathcal F}: {\mathcal C} \rightarrow
{\mathcal D}$ is the equivalence if, and only if \\ a)${\mathcal
F}$ is fully faithful, and \\ b) any $Y \in Ob \;{\mathcal D}$ is
isomorphic to an object of the form ${\mathcal F}(X)$ for an
object $X \in Ob \;{\mathcal C}$ .
\end{theorem}
\subsection{The Fourier-Mukai transform}
Let $A$ be an abelian variety and ${\hat A}$ the dual abelian
variety which is by definition a moduli space of line bundles of
degree zero on $A$\cite{Glazunov:mumford}.The {\em Poincar{\'e}
bundle} ${\mathcal P}$ is a line bundle of degree zero on the
product $A \times {\hat A}$, defined in such a way that for all $
a \in {\hat A}$ the restriction of ${\mathcal P}$ on $A \times
\{a\}$ is isomorphic to the line bundle corresponding to the point
$a \in {\hat A}.$ This line bundle is also called {\em the
universal bundle}. Let
\[
{\pi}_{A}: \; A \times {\hat A} \rightarrow A,
\]
\[
{\pi}_{\hat A}: \; {\hat A} \times  A \rightarrow {\hat A},
\]
and ${\mathcal C}_{A}$ be the category of $O_{A}-$modules over $A,
\; M \in Ob \; {\mathcal C}_{A},$
\[
  {\hat S}(M) = {\pi}_{{\hat A},*}({\mathcal P} \otimes
  {\pi}_{A}^{*} M).
\]
Then, by definition, {\em the Fourier-Mukai transform} ${\mathcal
F}{\mathcal M}$ is the derived functor $R{\hat S}$ of the functor
${\hat S}.$  Let ${\mathcal D}(A), \; {\mathcal D}({\hat A})$ be
bounded derived categories of coherent sheaves on $A$ and ${\hat
A}$ respectively.
\begin{theorem}(Mukai.)
 The derived functor ${\mathcal F}{\mathcal M} = R{\hat S}$
induces an equivalence of categories between two derived
categories ${\mathcal D}(A)$ and ${\mathcal D}({\hat A}).$
\end{theorem}

\section{Cutset method and its implementation}

A quiver can be interpreted as a directed graph. Here we consider
quivers without multiple edges and loops.
The solution of some problems requires to cut all cycles or
constract a spanning-tree of a directed graph.
Rooted graph $ G $ is the graph that each node of $ G $
is reachable from a node $ r \in G .$
By the {\it cutset} of a graph we shall understand an
appropriated subset of nodes (called cutpoints) such
that any cycle of the graph contains at least one
cutpoint. DFS is the abbreviation of Depth First Search
method. During DFS we numbering nodes and label (mark)
edges. By Tarjan~\cite{Glazunov:tarjan} the DFS method has linear
complexity.
  The main function Cutsetdg of the package CUTSETDG
computes Cutset (a subset of
vertices which cut all cycles in the graph) of arbitrary
rooted directed graph. It uses 3 functions: Adjarcn,
Cutpoints and Unicut. For this program the author developed
rather simple and efficient algorithm that based on
DFS-method. The function is the base for Cutset methods
for systems of procedures.
The package is implemented by the Allegro Common Lisp
language~\cite{Glazunov:guy&steele,Glazunov:CL}. \\
\begin{note}
It is not difficult to modify the
Cutset algorithm for processing directed graphs with multiple edges.
\end{note}
Package CUTSETDG \\
 Title: Cutset of a rooted directed graph   \\
 Summery:
 This package implements computation of a cutset
 of a rooted directed graph. The graph have to defined by adjacency
 list. \\
 Example of Call:   (Cutsetdg 'a)
 where "a" is a root of exploring graph. \\
{\bf The Common Lisp text of the package CUTSETDG} \\
\begin{verbatim}
;; Name: Cutsetdg
;;
;; Title: Cutset of a rooted directed graph
;;
;; Author: Nikolaj M. Glazunov
;;
;; Summery:
;; This package implements computation of a cutset
;; of a rooted directed graph. The graph have to defined by adjacency
;; list
;;
;; Example of Call:   (Cutsetdg 'a)
;; where "a" is a root of exploring graph
;;
;; Allegro CL 3.0.2
;;

(DEFUN Cutsetdg (startnode)

;; startnode is a root of exploring graph
;; DFS - Depth First Search method for connected graph
;; st - stack
;; v - exploring node
;; sv -  son of the exploring node
;; inl - inverse edges list
;; df -  DFS-numbering of the node (property)
;; Outarcs - adjacency list of Output arcs of the node (property)
;; 1, 2  - lables

    (SETF (GET startnode 'df ) 1)   ;DFS-numbering is equal 1
   (SETF (GET startnode 'Outarcs)
            (Adjarcn  (LIST startnode))) ;;startnode obtained the a-list
                                 ;; property Outarcn (Output arcs of the
                                 ;; node)
    (PROG ( st v sv inl  cuts)
         (PUSH startnode st)
   1   (SETQ v (CAR st)) ;;explored node received value from stack of nodes
   2      (COND ((NOT (EQ  NIL (GET v 'Outarcs)))
                    (SETQ sv (CAAR (GET v 'Outarcs)))
                        (COND ((EQ NIL (GET sv 'df))
                            (SETF (GET sv 'df) (+ 1 (GET v 'df)))
       ;; modification of the edge of son's adj-list
       (SETF (GET sv 'Outarcs)
            (Adjarcn  (LIST sv))) ;; node sv obtained the adj-list property
                                 ;; Outarcs (adjacency Outarcs)
          (SETF (GET v 'Outarcs) (REMOVE (LIST sv v) (GET v 'Outarcs)
                              :TEST 'EQUAL))
                           (PUSH sv st)
                            (GO 1)
            )
              (T
                 (COND ((AND (> (GET v 'df) (GET sv 'df))
                            (NOT (EQUAL v sv)))
              (SETQ inl (CONS (CAR (GET v 'Outarcs)) inl))
              (SETF (GET v 'Outarcs) (REMOVE (LIST sv v) (GET v 'Outarcs)
                              :TEST 'EQUAL))
                        (GO 2)
                        )
              (T
               (SETF (GET v 'Outarcs) (REMOVE (LIST sv v) (GET v 'Outarcs)
                              :TEST 'EQUAL))
                (GO 2)
                )
                       )
                )
                   )
                 )
                              (T
                (POP st)
         (COND     ((NOT (NULL st))
                      (GO 1)
              )
                     (T                 ;; end
                   (SETQ cuts (Unicut (Cutpoints inl)))
                (RETURN cuts)
                )
              )
                ))
     )
   )
(DEFUN Adjarcn (PATH)
      (MAPCAN #'(LAMBDA (E)
              (COND ((MEMBER (CAR E) (CDR E)) NIL)
                   (T (LIST E))))
          (MAPCAR #'(LAMBDA (E)
                            (CONS E PATH))
             (CAR  (GET (CAR PATH) 'NEIGHBORS)))))
(DEFUN Cutpoints (inl)
        (MAPCAR 'CAR inl))
(DEFUN Unicut (cpl)
     (COND ((NULL cpl) NIL)
           (T (CONS (CAR cpl)
                (Unicut (REMOVE (CAR cpl) cpl)))
             )
     )
     )
\end{verbatim}

\subsection*{Acknowledgements}

I would like to thanks the organizers of the conference
SymmNMPh'2003 and  NATO ASI "Computational Noncommutative Algebra
and Applications" and  for providing a very pleasant environment
during the conferences and NATO ASI for support.

\end{document}